\documentclass[twocolumn,aps,superscriptaddress]{revtex4-1}

\usepackage{eurosym}
\usepackage{amsfonts}
\usepackage{mathrsfs}
\usepackage{color}
\usepackage{xspace}
\usepackage{longtable}
\usepackage{xspace}
\usepackage{multirow}
\usepackage{tabularx}
\usepackage{amsmath}
\usepackage{amssymb}
\usepackage{graphicx}
\usepackage{dcolumn}
\usepackage{bm}
\usepackage[colorlinks,urlcolor=blue,citecolor=blue,linkcolor=blue]{hyperref}
\usepackage[normalem]{ulem}
\usepackage{mathrsfs}
\usepackage{epstopdf}
\usepackage{appendix}
\usepackage[level]{datetime}
\usepackage{ulem}
\usepackage{verbatimbox}

\begin{document}

\title{Kibble-Zurek behavior in a topological phase transition with a quadratic band crossing }
\date{\today}

\author{Huan Yuan}
\author{Jinyi Zhang}
\author{Shuai Chen}
\affiliation{Hefei National Research Center for Physical Sciences at the Microscale and School of Physical Sciences, University of Science and Technology of China, Hefei 230026, China}
\affiliation{Shanghai Research Center for Quantum Sciences and CAS Center for Excellence in Quantum Information and Quantum Physics, University of Science and Technology of China, Shanghai 201315, China}
\affiliation{Hefei National Laboratory, Hefei 230088, China}
\author{Xiaotian Nie}
\email{nxtnxt@hfnl.cn}
\affiliation{Hefei National Laboratory, Hefei 230088, China}
\date{\today }

\begin{abstract}
Kibble-Zurek (KZ) mechanism describes the scaling behavior when driving a system across a continuous symmetry-breaking transition. Previous studies have shown that the KZ-like scaling behavior also lies in the topological transitions in the Qi-Wu-Zhang model (2D) and the Su-Schrieffer-Heeger model (1D), although symmetry breaking does not exist there.
Both models with linear band crossings give that $\nu \simeq1$ and $z \simeq1$. A natural question is whether different critical exponents can emerge in topological transitions beyond the linear band crossing.
In this work, we look into the KZ behavior in a topological 2D checkerboard lattice with a quadratic band crossing. We investigate from dual perspectives: momentum distribution of the Berry curvature in clean systems for simplicity, and real-space analysis of domain-like local Chern marker configurations in disordered systems, which is a more intuitive analog to conventional KZ description.
In equilibrium, we find the correlation length diverges at the transition with a power $\nu\simeq 1/2$. Then, by slowly quenching the system across the topological phase transition, we find that the freeze-out time $t_\mathrm{f}$ and the unfrozen length scale $\xi(t_\mathrm{f})$ both satisfy the KZ scaling, verifying $z\simeq 2$.
We subsequently explore KZ behavior in topological phase transitions with other higher-order band crossings and find the relationship between the critical exponents and the order. Our results extend the understanding of the KZ mechanism and non-equilibrium topological phase transitions.
\end{abstract}

\maketitle


\section{Introduction}
Quantum phase transitions are one of the fascinating topics in modern physics. Conventional continuous phase transitions are driven by spontaneous symmetry breaking and are usually described in terms of local order parameters that take continuous values \cite{book_SymBre,Matthias2003,Sachdev_2011,Nie202303,Jara2024}. In contrast, topological phase transitions are a new class of transitions beyond Landau's symmetry breaking theory, characterized by quantized, non-local topological invariants \cite{Hasan2010,Shou-Cheng2011}.
Recently, the non-equilibrium dynamics of topological phase transitions have received widespread attention both experimentally and theoretically \cite{Weitenberg2018,Alessio2015,Hauke2014,Fuxiang2020,Sun2018,Nie202310}, particularly concerning the critical behavior of quenching across topological phase transitions \cite{Liou2018,Toma2019,Toma2020,Fuxiang2022,Kuo2021,deng2024}.

Relating conventional continuous phase transitions and nonequilibrium dynamics, the Kibble-Zurek (KZ) mechanism was originally proposed by Kibble in his study of the formation of topological defects in the early universe\cite{Kibble1976}, and it has been applied by Zurek to study phase transitions in condensed matter systems \cite{Zurek1985,Zurek2005}. It elucidates the universal scaling of a system when it is quenched across a continuous phase transition in a finite time $\tau$. In equilibrium, the theory of critical phenomena states that as the control parameters $V$ are tuned closer to their critical values $V_\mathrm{c}$, the correlation length $\xi$ and relaxation time $t_\mathrm{r}$ tend to diverge as $t_\mathrm{r}\sim|V-V_\mathrm{c}|^{-z\nu}$ and $\xi\sim|V-V_\mathrm{c}|^{-\nu}$, where $z$ is the dynamic exponent and $\nu$ is the correlation length critical exponent. As the system slowly quenches across the transition, the relaxation time diverges as it approaches the critical point, and the system's evolution ceases to be adiabatic. The time interval during which the system no longer evolves adiabatically is characterized as the freeze-out time $t_\mathrm{f}$. The freeze-out time and the unfrozen correlation length satisfy $t_\mathrm{f}\sim \tau^{z\nu/(1+z\nu)}$ and $\xi(t_\mathrm{f})\sim \tau^{\nu/(1+z\nu)} $.
The KZ behavior has been observed in diverse experimental systems, such as cosmic microwave background \cite{Bevis2008}, superfluids \cite{Chesler2015,Mathey2010}, superconductor \cite{Sonner2015,Garaud2014,Monaco2006}, liquid crystals \cite{Isaac1991}, and ultracold atoms \cite{Esslinger2007,Weiler2008,Shin2019,Chen2011,Lamporesi2013,Navon2015,Clark2016,Lukin2019,Yi2020,Jiazhong2021,LIANG2022}.

Although topological transitions can not be described by Landau's symmetry breaking theory, the KZ-like behavior has already been introduced to topological systems. Some studies have manifested that the total density of excitations shows a KZ power-law scaling \cite{Liou2018,Toma2019,Bermudez2009,Bermudez2010,Sim2022,Sim2023}. However, this simple generalization does not reveal the underlying fundamental topological physics and shows incapability in extracting correlation lengths. 
Later, it is afterwards proposed that effective correlation length can be extracted from the Berry curvature profile in momentum space, and it shows power-law divergence at the transition \cite{Wei2017}. Yet, this momentum-space description is not quite intuitive in contrast with the conventional KZ mechanism.
Recently, significant breakthroughs have been made in the real-space analysis.
It has been found that the local Chern marker (LCM) will show domain-like configurations when the system is disordered \cite{Toma2020}. As long as the disorder is weak, it is shown that the disorder-averaged normalized autocorrelation function of LCM  can be regarded as an intrinsic property of the original clean topological system. Thus, the correlation length can be determined via this purely real-space method. During the quench, the domain-like configurations are the analogy to conventional topological defects and the correlation length shows KZ behavior. This method is first proposed in two-dimensional Qi-Wu-Zhang (QWZ) insulators \cite{Toma2020} and then extended to one-dimensional Su-Schrieffer-Heeger (SSH) models \cite{Fuxiang2022} by changing LCM to local winding number. However, since these two models have similar linear dispersion during gap closing, they have yielded the same critical exponents, $\nu \simeq1, z \simeq1$. Therefore, we naturally ask whether different critical exponents can be obtained.

In this work, we systematically investigate the KZ behavior in the topological transition of a 2D checkerboard lattice model with a quadratic band crossing. For every scaling studied, dual perspectives are adopted: momentum-space Berry curvature in the clean system, and the real-space LCM in the disordered system. In the former, we mainly focus on the radial distribution of Berry curvature. In the latter, we are concerned with the domain-like configurations of the LCM and extract the real-space correlation lengths.
\begin{figure}[t]
	\centering
	\includegraphics[width=1\linewidth]{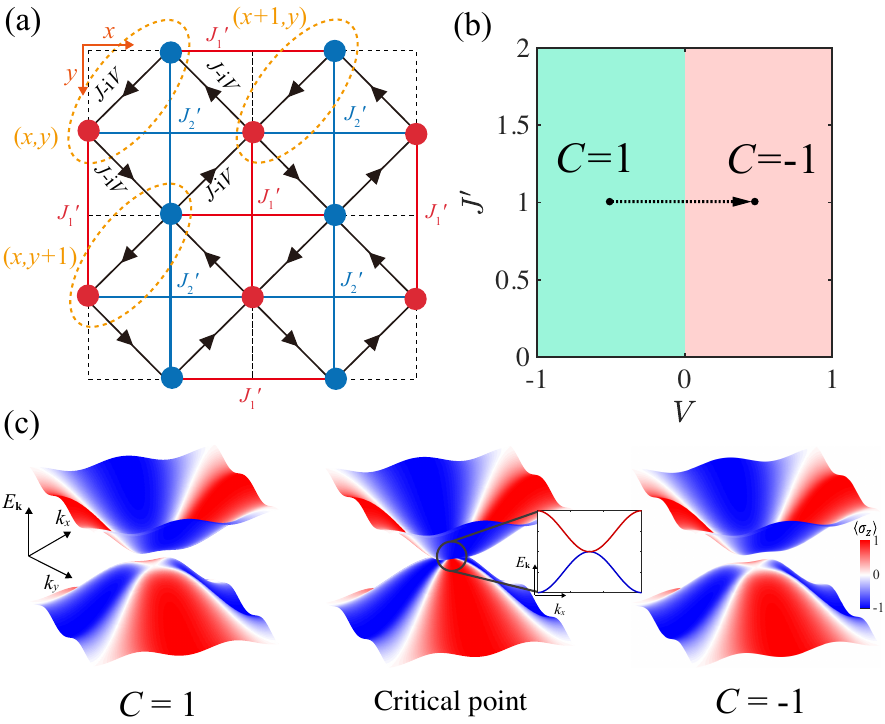}
	\caption{(a). Checkerboard lattice model. Red (Blue) points denote the A (B) sublattices. NN hopping strength along the black arrows direction (opposite direction) is $J-\mathrm{i}V$ ($J+\mathrm{i}V$), which breaks the time-reversal symmetry as long as $V\neq 0$. Red (Blue) solid lines represent the NNN hopping strength $J_1^{\prime}(J_2^{\prime})$. (b). Topological phase diagram. Chern number $|C|=1$ acquires an opposite sign to $V$. At the transition, it is a time-reversal invariant semimetal. (c). The band structure of our model is in different regimes. The colors denote the polarization $\langle\sigma_z\rangle$. The gap closes with a quadratic dispersion at the transition $V=0$. We can see the band-inversion surface (BIS, shown as the white contours) is not a circle around symmetric momentum as in the QWZ model \cite{Zhang2018,Sun2018}, which is a fundamental effect induced by the quadratic band crossing.} \label{Setup_Band}
\end{figure}
We find that the length scale, in both perspectives, diverges near the critical point of the topological phase transition in equilibrium, satisfying $\xi\sim|V-V_\mathrm{c}|^{-1/2}$. By slowly quenching through the topological phase transition over a quench time $\tau$, we observe that the freeze-out time $t_\mathrm{f}$, defined by the delay time of the relaxation of Hall response in the clean system or the divergence of correlation length in the disordered system, satisfies $t_\mathrm{f}\sim \tau^{1/2} $. The corresponding length scale at the freeze-out time $t_\mathrm{f}$ satisfies $\xi(t_\mathrm{f})\sim \tau^{1/4}$, where the dynamic exponent $z \simeq2$ and correlation length critical exponent $\nu \simeq1/2$ differs from the results of linear band dispersion.
Additionally, we extend the analysis to higher-order dispersions beyond the quadratic band crossing and find that the KZ power-law relations still hold. The corresponding critical exponents are related to the band-crossing dispersion.

\section{Model}
Two-dimensional (2D) checkerboard lattice \cite{Kai2009,Kai2011} is an example of a topological model with a quadratic band crossing. We consider a simplified checkerboard lattice model consisting of two sublattices (A and B) with real and isotropic nearest-neighbor (NN) hoppings ($J$), two kinds of next-nearest-neighbor (NNN) hoppings ($J_1^{\prime},J_2^{\prime}$), and complex NN hopping ($\mathrm{i}V$) leading to a topological nontrivial phase (see Fig.\ref{Setup_Band}(a)). The Hamiltonian can be written as:

\begin{align}
	\begin{split}
			\hat{H}&= \sum_{\mathbf{r}}|x, y\rangle\langle x, y| \otimes\begin{pmatrix} 0& J-\mathrm{i}V \\ J+\mathrm{i}V & 0 \end{pmatrix} \\
			&+\sum_{\mathbf{r}}\left[|x, y\rangle\langle x+1, y| \otimes \begin{pmatrix} J_2^{\prime}& 0 \\ J- \mathrm{i}V & J_1^{\prime} \end{pmatrix} +\text{H.c.}\right]\\
			&+\sum_{\mathbf{r}}\left[|x, y\rangle\langle x, y+1| \otimes \begin{pmatrix} J_1^{\prime}& J+ \mathrm{i}V \\ 0 & J_2^{\prime} \end{pmatrix} +\text{H.c.}\right]\\
			&+\sum_{\mathbf{r}}\left[|x, y+1\rangle\langle x+1, y| \otimes \begin{pmatrix} 0& 0 \\ J+\mathrm{i}V & 0 \end{pmatrix} +\text{H.c.} \right],\\
	\end{split}
\end{align}
where $\mathbf{r}=(x,y)$ denote the unit cell position (a unit cell is shown as a dashed ellipse in Fig.\ref{Setup_Band}(a)), and the $2\times 2 $ matrices represent the sublattice tunneling related to the unit cell hopping.


After a Fourier transformation, we can express the Hamiltonian in momentum space
$\hat{\mathcal{H}}_{\mathbf{k}}$ with momentum $\mathbf{k}=(k_x,k_y)$ as
\begin{align}
	\begin{split}
		\label{Hk}
	 	\hat{\mathcal{H}}_{\mathbf{k}}= & \left[-\left(J_1^{\prime}+J_2^{\prime}\right)\left(\cos k_x+\cos k_y\right) \right] \hat{I} \\
	 	&-4 J \left(\cos \frac{k_x}{2} \cos \frac{k_y}{2}\right) \hat{\sigma}_x  -2V \left(\sin \frac{k_x}{2} \sin \frac{k_y}{2}\right) \hat{\sigma}_y \\
	 	&-\left(J_1^{\prime}-J_2^{\prime}\right)\left(\cos k_x-\cos k_y\right) \hat{\sigma}_z,
	\end{split}
\end{align}
where $\hat{I}$ and $\hat{\sigma}_{x,y,z}$ are the identity and Pauli matrices, respectively. We set $J=1$, $J_1^{\prime}=-J_2^{\prime}=J^{\prime}$ and periodic boundary condition for the entire text afterwards. It is easy to read from Eq.\ref{Hk} that the system is an insulator without time-reversal symmetry as long as $V\neq 0$. Actually, the system has two topologically non-trivial phases, whose Chern number $C$ is determined by the sign of complex hopping $V$, as shown in Fig.\ref{Setup_Band}(b). Additionally, the energy band structures in different regimes are depicted in Fig.\ref{Setup_Band}(c). We can see that the energy gap closes at $\mathbf{G} = (k_x, k_y) = (\pi, \pi)$ when $V=V_\mathrm{c}=0$, exhibiting a quadratic band crossing with Chern number changing by $2$. This is totally different from the QWZ model. There, the gap closes at a Dirac cone, and every Dirac cone changes the Chern number with $1$. Besides, the band-inversion surface (BIS, defined as the $\langle\sigma_z\rangle=0$ contour) in the checkerboard lattice model forms a crisscross, while in the QWZ model, it is a circle around the band crossing point \cite{Zhang2018,Sun2018}.

\section{Ground state}
For conventional second-order phase transitions driven by spontaneous symmetry breaking, the correlation lengths exhibit a universal scaling behavior as the systems approach the critical point. 
When it comes to topological transitions, such as the QWZ model and the SSH model, a scaling of $\nu \simeq1$ has been found. However, can this exponent be different in other models with different band crossings?

In the following, we investigate the checkerboard model from two perspectives: the momentum space in the clean system, and the real space in the disordered system.

\subsection{Perspective I: Clean system}
\begin{figure}[t]
	\centering
	\includegraphics[width=\linewidth]{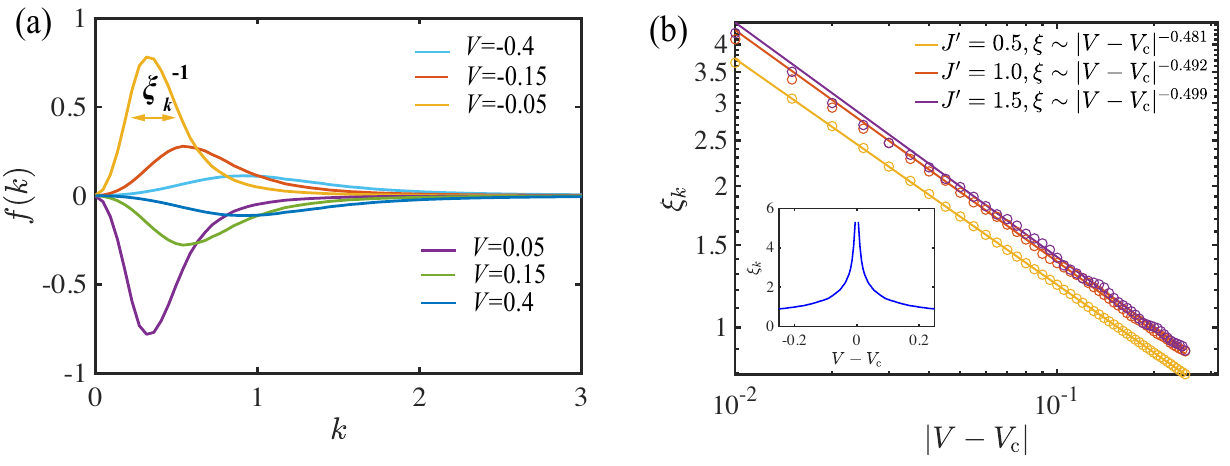}
	\caption{(a). Correlation length in the ground state could be determined by the FWHM of the radial distribution of Berry curvature, here $J^{\prime}=1$. The closer to the transition, the narrower the peak. (b). This correlation length diverges in a power $\nu\simeq 1/2$ as the system approaches the phase transition.}
	\label{Kspace_CorLen_Equi}
\end{figure}
Some studies have provided correlation functions for topological phases, defined as the Fourier transform of the Berry curvature \cite{Wei2017,Wei2019,ChenWei2023sci}, which measures the itinerant-circulation correlation between Wannier states \cite{Thonhauser2005,Marzari2012}. Alternatively, a more direct definition of correlation length can also be extracted from the width of the profile of the Berry curvature, which will be explained in the following. Similar procedures have been used in studying the critical behavior in the QWZ insulator and the SSH model \cite{Toma2020,Fuxiang2022}. 

The valence band Berry curvature $F(\mathbf{k})$ can be expressed via Berry connection $A_{k_\alpha}=\left\langle u_{\mathbf{k}}\left|\mathrm{i} \partial_{k_\alpha}\right| u_{\mathbf{k}}\right\rangle$ ($\alpha=x,y$) as
\begin{align}\label{bc-eig}
F(\mathbf{k}) = \partial_{k_x} A_{k_y}-\partial_{k_y} A_{k_x},
\end{align}
where $\left|u_{\mathbf{k}}\right\rangle$ is the valence band Bloch wavefunction at momentum $\mathbf{k}$. The Chern number mentioned before can be calculated by $C=\frac{1}{2\pi}\int_{\rm{FBZ}} F(\mathbf{k}) \rm{d}^2 \mathbf{k}$, where "FBZ" represents the first Brillouin zone.

Then, by integrating the Berry curvature angularly, $f(k)=\frac{1}{k}\int_{|\mathbf{k}'-\mathbf{G}|=k}F(\mathbf{k}') |\text{d} \mathbf{k}' |$, we obtain its radial distribution $f(k)$.
Fig.\ref{Kspace_CorLen_Equi}(a) illustrates $f(k)$ with fixed $J^{\prime}= 1$ and six different values of complex NN hoppings strength $V$. It is evident that the radial distribution of the Berry curvature exhibits a peak. Initially, this peak width decreases with the increasing $V$, reaching its minimum at the topological phase transition point $V_\mathrm{c} = 0$, and subsequently increases. This reflects the critical behavior near the topological phase transition point.

To describe this critical behavior quantitatively, we define the correlation length $\xi_k$ by the inverse of the full width at half maximum (FWHM) value of the radial distribution of Berry curvature $f(k)$. A similar approach has been utilized in the QWZ insulators and the SSH model \cite{Toma2020,Fuxiang2022}. As shown in Fig.\ref{Kspace_CorLen_Equi}(b), we observe that as the system approaches the critical point $V_\mathrm{c}  = 0$, the correlation length diverges. We plot the relationship between the correlation length and $|V - V_\mathrm{c}|$ in log-log coordinates under three sets of parameters, namely $J_1^{\prime}=-J_2^{\prime}=J^{\prime} = 0.5$, $1$, $1.5$. We find that the correlation length satisfies $\xi_k \sim |V - V_\mathrm{c}|^{-\nu}$, where $\nu$ is approximately $1/2$. This critical exponent value is distinct from the case of linear dispersion, such as in the QWZ insulators where $\nu \simeq 1$.

\begin{figure*}[hbt]
	\centering
	\includegraphics[width=0.8\linewidth]{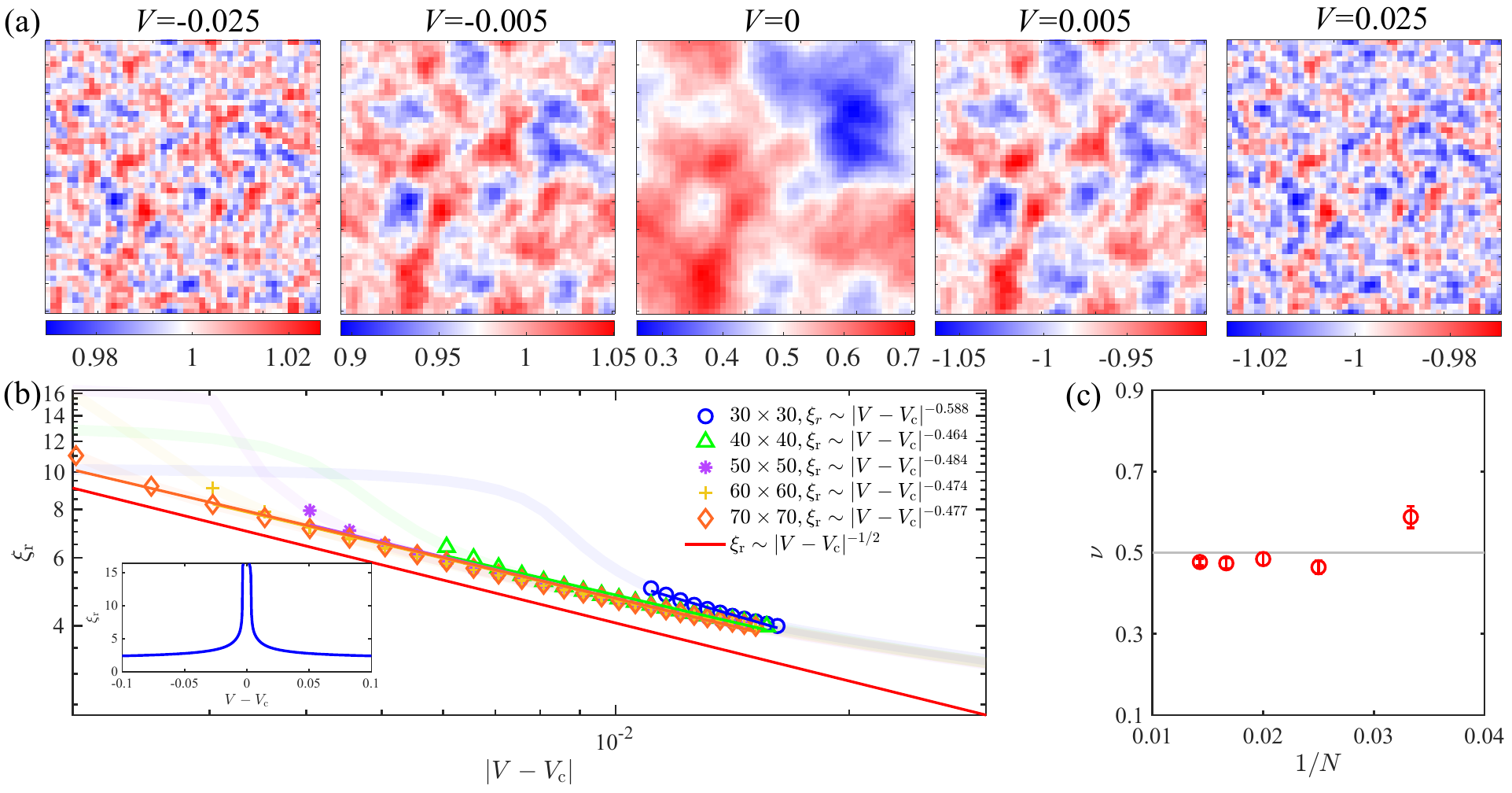}
	\caption{(a). The real-space distribution of the LCM in a disordered sample with size $N\times N=50\times 50$. The area of the fragment domain increases to the order of the whole system sizes at the transition. (b). Divergence of real-space correlation length $\xi_r$ at transition with different system sizes in a log-log scale. Each data is an average of $20$ disorder samples. For a certain system size $N$, the correlation length shows good power law in $4<\xi<N/6$ regime (emphasized by the markers). The red line represent a $\left|V-V_{\mathrm{c}}\right|^{-1/2}$ scaling for reference. (c). The fitted power $\nu$ approaches $1/2$ as the system size increases. Here, the NNN hopping strength is set to $J^{\prime}=1$ and the disorder strength to $\delta V = 0.05$.}
	\label{Rspace_CorLen_Equi} 
\end{figure*}
\subsection{Perspective II: Disordered system}

Then, we switch to the other perspective: real-space analysis in disordered systems. Following the previous studies, correlation length can be characterized by the real-space distribution of the local Chern marker (LCM) when the system is weakly disordered, where the LCM $c(\mathbf{r})$ is calculated via \cite{Prodan2010,Bianco2011,Toma2020}
\begin{eqnarray}
	\label{LCMoperator}
	c(\mathbf{r}) = 2\pi \mathrm{i} \mathrm{Tr}\left\{\hat{P}(\mathbf{r}) \hat{P}_\mathrm{val}\left[-\mathrm{i}\left[\hat{x},\hat{P}_\mathrm{val}\right],-\mathrm{i}\left[\hat{y},\hat{P}_\mathrm{val}\right]\right]\right\}.
\end{eqnarray}
Here $\hat{P}(\mathbf{r})$ is the local projector to the unit cell at $\mathbf{r}$, $\hat{P}_\mathrm{val}$ is the projector to the valence band (the lower half spectrum), and $\hat{x},\hat{y}$ are the position operators. 

If the system is clean, the LCM is uniform and equal to the Chern number in the ground state.
While in a disordered system,
as long as the disorder is weak, the mean value of LCM of the ground state is still the Chern number of its original clean counterpart in the thermodynamic limit \cite{ChenWei2024}. In a specific disorder sample, the local deviation from the mean value shows a domain-like configuration (see Fig.\ref{Rspace_CorLen_Equi}(a)), whose amplitude is proportional to the disorder strength, but its shape is independent \cite{Toma2020}. Besides, if there are enough disorder samples, the disorder-averaged normalized autocorrelation function of LCM  converges to a certain limit (see Appendix for more details). 
Therefore, the disorder-averaged normalized autocorrelation function calculated from a weakly disordered system can be regarded as an intrinsic character of its clean counterpart, and the correlation length determined by this method is a intrinsic quantity of the topological system.

We consider the weak disorder is added on the time-reversal-breaking hopping term, $V\to V_\mathrm{dis} (\mathbf{r})$, with an independent uniform random distribution $V_\mathrm{dis} (\mathbf{r})\in [V-\delta V, V+\delta V]$ with $\delta V = 0.05$. The real-space distribution of the LCM in a specific disordered sample is illustrated in Fig.\ref{Rspace_CorLen_Equi}(a) for several values of $V$, ranging from topological regions with $C=1$, across the critical point to another topological phase with $C=-1$. At the transition, the sizes of domains reach their maxima for about the whole system size, which indicates the divergence of correlation length in the thermodynamic limit.

We identify the correlation length in the real space $\xi_r$ to the distance where the disorder-averaged autocorrelation function of the LCM distribution drops to zero \cite{Toma2020,Fuxiang2022,ChenWei2023sci}. Fig.\ref{Rspace_CorLen_Equi}(b) shows that $\xi_r$ exhibits power-law scaling as $V$ approaches the critical point in the intermediate correlation length regime, i.e. $4<\xi_r<N/6$. That is because we are dealing with a finite-size lattice model, the maximal $\xi_r$ can not exceed the system size (infrared limit), and the minimal $\xi_r$ must saturate to the order of one lattice constant (ultraviolet limit). Only the intermediate regime suffers less from these two limits.

As the system size increases, the estimated scaling exponent $\nu$ approaches $1/2$, which is consistent with the results obtained from the Berry curvature in the clean system. From the discussion above, we find that in a 2D checkerboard model with a
quadratic band crossing, the critical phenomena of the topological phase transition are present in both momentum space and real space, with a power-law exponent equal to half of the results of linear band crossings.

\section{Quench dynamics}
In this section, we investigate the KZ behavior of topological phase transitions in the 2D checkerboard lattice. The system is quenched across the topological transition by linearly changing the complex NN hoppings strength $V$ with time $t$, expressed as $V(t)=-1/2+t/\tau$, where $\tau$ denotes the quench time, which corresponds to the change in the Chern number from $1\left(t=0\right)$ to $-1\left(t=\tau\right)$ as shown by the arrow in the Fig.\ref{Setup_Band}(b).

We still work in two perspectives and focus on the freeze-out time $t_\mathrm{f}$ and the unfrozen correlation length $\xi(t_\mathrm{f})$.

\subsection{Perspective I: Clean system}

\subsubsection{Freeze-out time $t_\mathrm{f}$}
\begin{figure}[hbt]
	\centering
	\includegraphics[width=\linewidth]{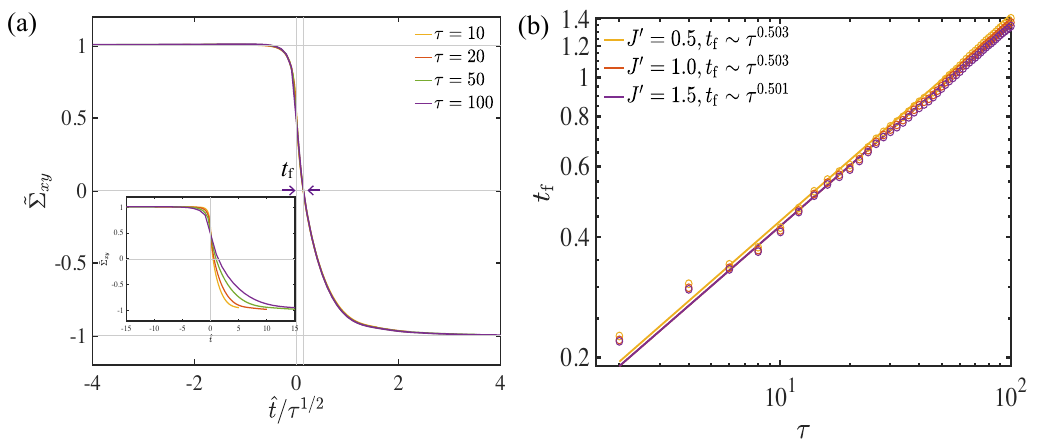}
	\caption{(a). The dynamics of non-equilibrium Hall response during the quench with $J^{\prime}=1$. These curves collapse in a rescaled time coordinate. We identify the freeze-out time $t_\mathrm{f}$ as the time when the Hall response crosses zero, $\tilde{\Sigma}_{xy}(\hat{t}=t_\mathrm{f})=0$. The inset shows curves in a non-rescale $\hat{t}$ coordinate. (b). Freeze-out time with different quench time in a log-log plot, $t_\mathrm{f}$ scales with approximately $\sim \tau^{1/2}$.}
	\label{Kspace_DelayTime_Quench}
\end{figure}
To explore the temporal scaling in the dynamical process, we choose to study the Hall response during the non-equilibrium process \cite{Kuo2021}. The non-equilibrium Hall response is given by the integral of the Berry curvature $F_\mathrm{b}(\mathbf{k},t)$ weighted by the occupation probability $P_\mathrm{b}(\mathbf{k},t)$ in all instantaneous bands \cite{Caio2016,Hu2016}:
\begin{align}
 \begin{split}
  \tilde{\Sigma}_{x y}(t)=\frac{1}{2 \pi} \sum_\mathrm{b} \int_{\mathrm{FBZ}} \text{d}^2 \mathbf{k}~ P_\mathrm{b}(\mathbf{k},t)F_\mathrm{b}(\mathbf{k},t).
 \end{split}
\end{align}
Here $\mathrm{b}\in \left\{\mathrm{val},\mathrm{con}\right\}$ refers to the instantaneous valence and conduction band. 
$P_\mathrm{b}(\mathbf{k},t)$ equals to the overlap between the eigenstates of the instantaneous Hamiltonian and the time-evolved wave function during the quench, while $F_\mathrm{b}(\mathbf{k},t)$ is calculated using the corresponding instantaneous eigenstates via Eq.\ref{bc-eig}.
We computed the evolution of the Hall response during the quench process, as depicted in Fig.\ref{Kspace_DelayTime_Quench}(a). The Hall response deviates from its original quantized value and relaxes to a new quantized value after the quench. By transforming the coordinates with $\hat{t}=t-t_\mathrm{c}$, we set the phase transition time $t_\mathrm{c}=\tau/2 $ as the zero point and the freeze-out time $t_\mathrm{f}$ is defined as the time at which the Hall response relaxes to the average of its initial and final values. We plot the relationship between the freeze-out time $t_\mathrm{f}$ and the quench time $\tau$ in the log-log scale and found that it follows a power-law relationship, with the corresponding exponent close to $1/2$. This matches the prediction of the KZ mechanism for linear quenching with $z\nu \simeq1$, $t_\mathrm{f}\sim \tau^{\frac{z\nu}{1+z\nu}}\simeq\tau^{1/2}$.
\subsubsection{Unfrozen correlation length $\xi_k(t_\mathrm{f})$}
\begin{figure}[!hbt]
	\centering
	\includegraphics[width=1\linewidth]{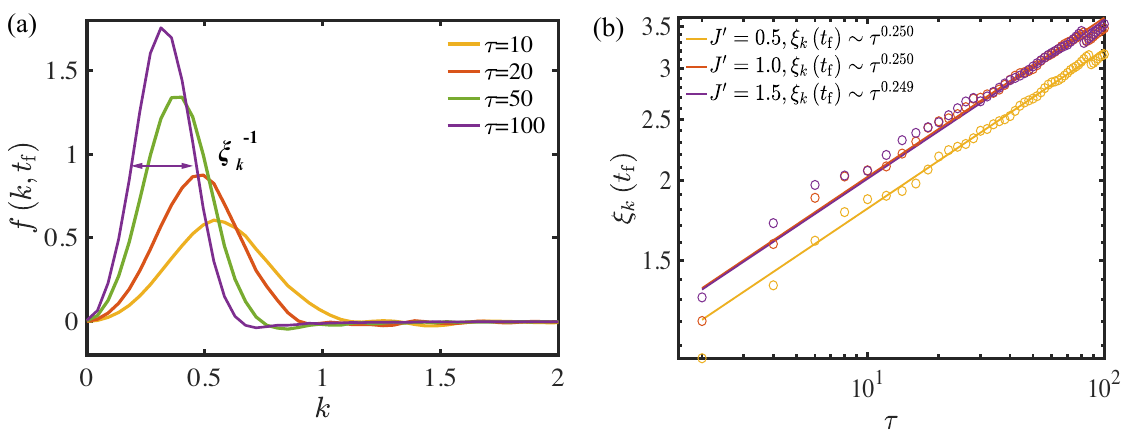}
	\caption{(a). Radial distribution of the Berry curvature at the freeze-out time $t_\mathrm{f}$ with $J^{\prime}=1$. (b). Correlation length $\xi_k$ at the freeze-out time obtained by the Berry curvature has a scaling with quench time $\tau$, $\xi_k(t_\mathrm{f}) \sim \tau^{1/4}$.}
	\label{Kspace_CorLen_Quench}
\end{figure}

We subsequently investigate the spatial scalings. The KZ mechanism predicts that at the freeze-out time $t_\mathrm{f}$, the correlation length $\xi_k(t_\mathrm{f})$ follows a power-law scaling with $\tau$. By obtaining the evolving wave functions during the quenching process, we calculated the corresponding Berry curvature and extracted the correlation length. Fig.\ref{Kspace_CorLen_Quench}(a) illustrates the radial distribution of the Berry curvature at corresponding freeze-out time $f(k,{t_\mathrm{f}})$ under different quench time. We observed that the width of the peak decreases with increasing quench time. Furthermore, the relationship between the extracted correlation length and the quench time is plotted for $J^{\prime}=0.5,1,1.5$. We found power-law relationships between the correlation length and quench time for different correlation lengths, with corresponding critical exponents very close to $1/4$. From the critical exponent of the correlation length at equilibrium, we obtain $\nu \simeq  1/2$. The power-law behavior of the freeze-out time gives the dynamic exponent $z \simeq  2$. According to the KZ mechanism, the unfrozen correlation length scales as $\xi(t_\mathrm{f}) \sim \tau^{\nu/(1+z\nu)} \simeq \tau^{1/4}$. Therefore, our results are consistent with the KZ prediction.
This critical exponent with a quadratic band crossing is half of the linear band crossing \cite{Toma2020,Fuxiang2022}.

\subsection{Perspective II: Disordered system}
\subsubsection{Freeze-out time $t_\mathrm{f}$}
When it comes to disordered systems, we need real-space analysis to show analogs to the formation of topological defects. Therefore, we switch back to explore the LCM distribution.

\begin{figure*}[!hbt]
	\centering
	\includegraphics[width=1\linewidth]{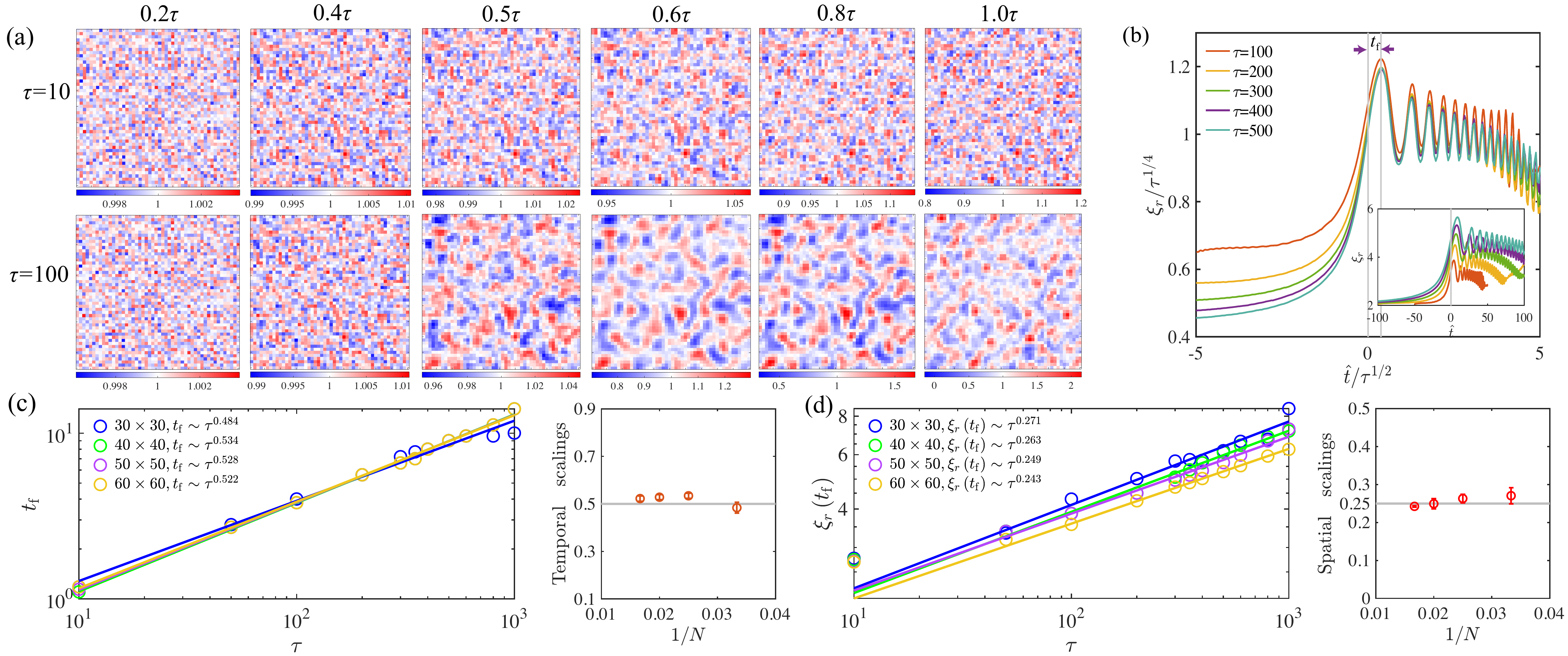}
	\caption{(a). The spatial distribution of the LCM during a fast ($\tau = 10$) and a slow ($\tau = 100$) quench with size $N=50$. The slower the quench, the larger the post-quench domain size. 
		(b). The dynamics of correlation length in real space in a double-rescaled coordinate. Curves collapse near the transition point. The freeze-out time $t_\mathrm{f}$ can be determined by the time delay of the trend of divergence in $\xi_r$ while crossing the transition. The inset shows the same data in a non-rescaled coordinate. Each data is an average from $10$ disorder samples. (c). Left panel: freeze-out time has a power-law scaling with the quench time, and the power-law exponent is close to $1/2$ for different system sizes. Right panel: estimated exponents with different system sizes.
		(d). Left panel: correlation length at the freeze-out time shows a scaling behavior, and the power-law exponent is close to $1/4$ for different system sizes. Right panel: estimated exponents with different system sizes. Here, the NNN hopping strength is set to $J^{\prime}=1$ and the disorder strength to $\delta V = 0.05$.} 
	\label{Rspace_Freezeout_time_CorLen_Quench}
\end{figure*}

The distribution of the LCM in quench dynamics can be calculated by replacing the projector onto the valence band $\hat{P}_\mathrm{val}$ with the projector onto the instantaneous occupied subspace. 
The evolution of the LCM profile during the quench process with $\tau=10$ and $\tau=100$ under weak disorder is depicted in Fig.\ref{Rspace_Freezeout_time_CorLen_Quench}(a). 
At the beginning of the slow quench, their LCM profiles are similar since they both follow the adiabatic evolution. For longer quench time, the system enters into the freeze-out regime later and holds a larger domain size after quench.

Like before, we extract the evolution of the correlation length $\xi_r$ over different quenching time, as shown in Fig.\ref{Rspace_Freezeout_time_CorLen_Quench}(b). During the evolution, the correlation length $\xi_r$ adiabatically follows the instantaneous ground-state value initially. Near the transition, the system enters into the freeze-out regime. Although the correlation length still increases, the trend of divergence occurs later than the critical point, so it shows a maximum after the transition. Here, we claim this time delay could be identified as the freeze-out time $t_\mathrm{f}$ (we verify that this method is also applicable to momentum-space analysis in clean systems as discussed in Appendix). After that, the correlation length is stabilized around a constant larger than the ground-state value.

In Fig.\ref{Rspace_Freezeout_time_CorLen_Quench}(c), we observe that the freeze-out time $t_\mathrm{f}$ exhibits a power-law relationship with the quench time $\tau$, with approximately the same power-law exponent of $1/2$ as observed in the clean system. 

\subsubsection{Unfrozen correlation length $\xi_r(t_\mathrm{f})$}

Similarly, we extract the unfrozen correlation length $\xi_r(t_\mathrm{f})$ from the LCM at the freeze-out time $t_\mathrm{f}$ to analyze. We plot the relationship between unfrozen correlation length $\xi_r(t_\mathrm{f})$ and the quench time on a log-log scale in the Fig.\ref{Rspace_Freezeout_time_CorLen_Quench}(d). We find that $\xi_r(t_\mathrm{f})$ scales with $\tau$ in an exponent of approximately $1/4$. Correspondingly, the critical exponent of the correlation length at equilibrium yields $\nu \simeq 1/2$, and the power-law behavior of the freeze-out time gives the dynamic exponent $z \simeq  2$. According to KZ mechanism, the unfrozen correlation length is expected to scale as $\xi(t_\mathrm{f}) \sim \tau^{\nu/(1+z\nu)} \simeq  \tau^{1/4}$. Thus, our findings are in agreement with the KZ prediction as observed in the clean system. 

Therefore, in the checkerboard lattice with a quadratic band crossing, when quenching across the topological phase transition, we observe temporal and spatial power-law behavior analogous to the Kibble-Zurek (KZ) mechanism in both momentum space and real space. The temporal power-law exponent is the same as that of the linear band crossing, while the spatial power-law exponent is half of that.

\section{Outlook on higher-order band crossings}
The critical exponent obtained from the quadratic band crossing differs from that of the previously studied linear band crossings. Therefore, it remains uncertain whether the KZ power-law scaling behaviors persist in systems with higher-order band crossings. If so, the corresponding critical exponent and its relationship with the band dispersion need further investigation. Accordingly, we can write down a general low-energy two-band effective model with higher-order crossing in momentum space, and explore the KZ behavior via methods mentioned in clean systems.

The low-energy effective Hamiltonian reads \cite{Wei2017}
\begin{align}
	\begin{split}
		\hat{H}_p(k, \phi) & =\eta_p k^p \cos (p \phi) \hat{\sigma}_x+\eta_p k^p \sin (p \phi) \hat{\sigma}_y+M \hat{\sigma}_z,
	\end{split}
\end{align}
where $k$ and $\phi$ represent the distance and polar angle relative to band-crossing momentum, respectively and integer $p$ denotes its crossing order, with $p=1$ representing the linear band crossing and $p=2$ representing the quadratic band crossing, and so forth. The topological phase transition occurs at $M=0$ with the change in Chern number between before and after the transition given by $C(M > 0)-C(M< 0)=p$. We should clarify that this effective description is only valid around the band-closing momentum near the transition. As we are studying the slow quench physics close to the transition, the distinction between this and the original lattice model is not important here.

Based on this Hamiltonian, we can first compute the Berry curvature in the ground state and obtain the correlation length. We numerically verify that the correlation length in ground state satisfies $\xi_k\sim |M - M_{\mathrm{c}}|^{-\nu}$, where $\nu$ is approximately $1/p$.

Similarly to the previous analysis, we quench across the topological phase transition, measure the Hall response, and extract the power-law relationship between the freeze-out time and the quench time. We find that temporal scaling is independent of the order of the band crossing with the corresponding exponent close to $1/2$. Similarly, by measuring the time evolution of the Berry curvature, we obtain the correlation length at the freeze-out time under different quench time, which also satisfies a power-law relationship. We observe that the corresponding exponents are distinct and close to $(2p)^{-1}$.
\begin{figure}[hbt]
	\centering
	\includegraphics[width=\linewidth]{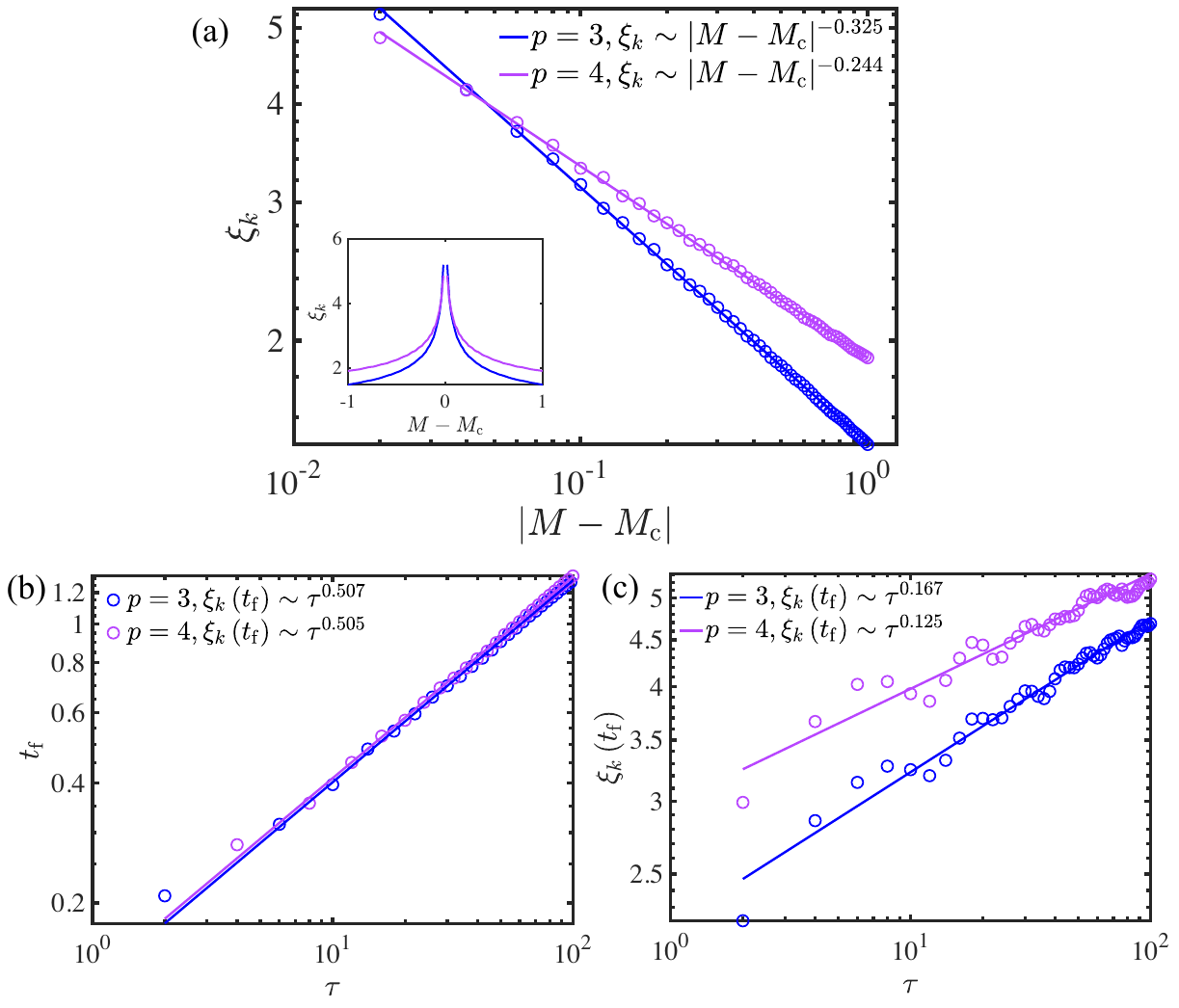}
	\caption{(a). Correlation length in the ground state diverges in a power $\nu \simeq 1/3$ for $p=3$ and $\nu \simeq 1/4$ for $p=4$ as the system approaches the phase transition.
	(b). Freeze-out time has a power-law scaling with the quench time, $t_\mathrm{f}\sim \tau^{1/2}$ with $p=3$ and $p=4$.
	(c). Correlation length at the freeze-out time shows a scaling behavior for $p=3$, $\xi_k(t_\mathrm{f})\sim \tau^{1/6}$ and $p=4$, $\xi_k(t_\mathrm{f})\sim \tau^{1/8}$.}
	\label{Kspace_high_order_time_delay_corlen_quench}
\end{figure}

\section{Summary}
In this work, we investigated the KZ behavior in a topological phase transition with a quadratic band crossing. We used dual perspectives: momentum space in clean systems and real space in disordered systems. Both perspectives give consistent scalings, which are distinct from previously studied linear band crossing models.
Further, we go beyond quadratic to higher orders. They also follow KZ behavior but acquire different critical exponents.

Our study extends the KZ behavior in topological transitions beyond the linear band crossing and confirms the consistency between correlation length from Berry curvature and LCM. For future studies, we get some hint from this work that the LCM correlation length may characterize the Anderson insulator transition (crossover) if the disorder is strong.

\begin{acknowledgments}
XN thanks Hui Zhai, Wei Zheng, Fan Yang and Hanteng Wang for their helpful discussions. This work is supported by NSFC (Grants No. GG2030007011 and No. GG2030040453) and Innovation Program for Quantum Science and Technology (Grants No. 2021ZD0302004).
The numerical calculations in this paper have been done on the supercomputing system in the Supercomputing Center of University of Science and Technology of China. This research was also supported by the advanced computing resources provided by the Supercomputing Center of the USTC.
\end{acknowledgments}

\appendixtitleon
\begin{appendices}

\section{Extraction of correlation length from the local Chern marker}

\begin{figure}[!htb]
	\centering
	\includegraphics[width=1\linewidth]{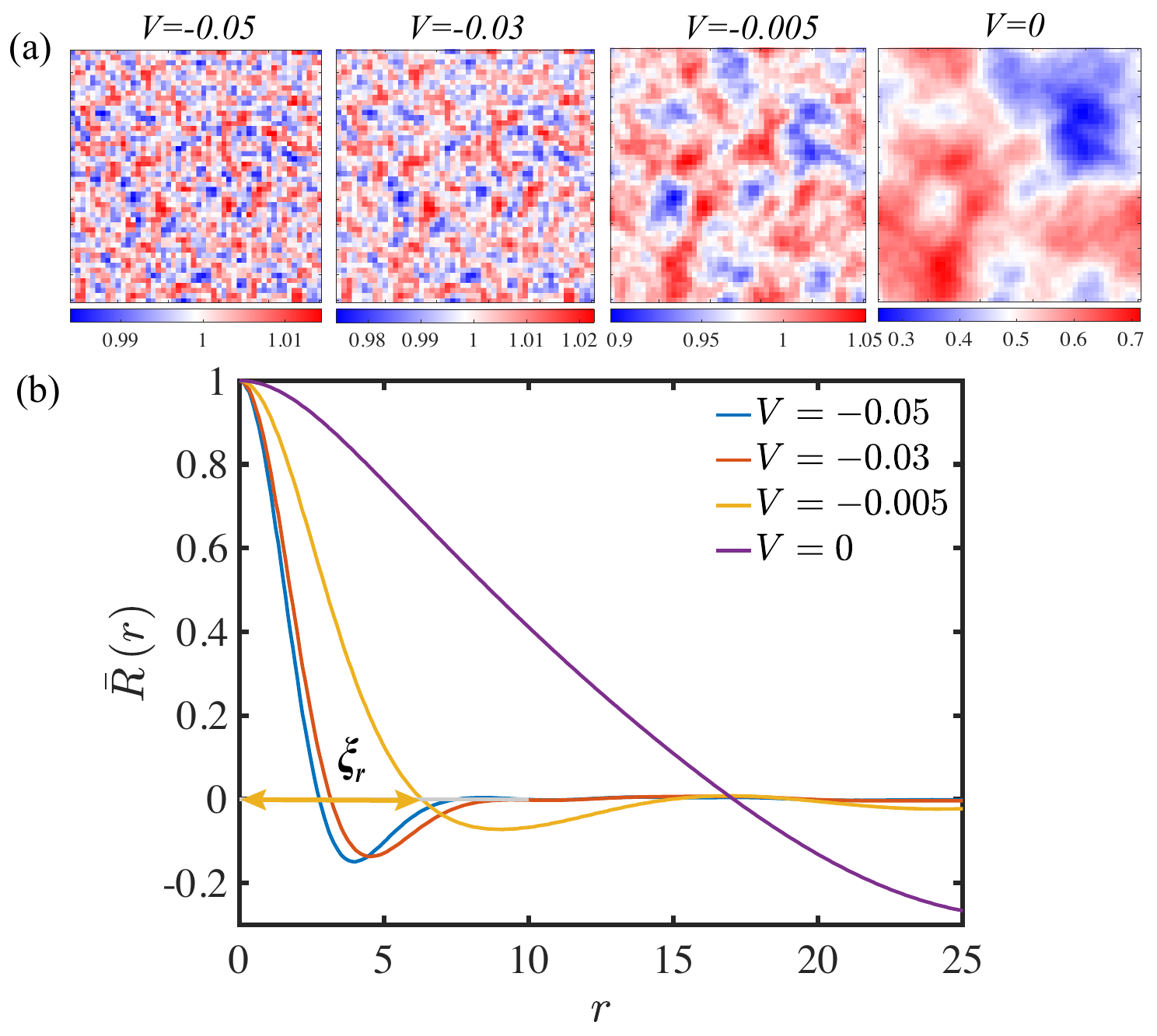}
	\caption{(a). The distribution of the LCM in ground state. (b). Disorder-averaged autocorrelation function calculated from the LCM. Here, the NNN hopping strength is set to $J^{\prime}=1$, and the average is performed over $20$ disorder realizations $\delta V = 0.05$ and size $N=50$. }
	\label{SM_Rspace_corfun}
\end{figure}

Fig.\ref{SM_Rspace_corfun}(a) shows the real-space distribution of the local Chern marker (LCM) in ground state. To estimate the size of inhomogenous "domains" in the LCM, we calculated the normalized autocorrelation function of the real-space distribution function $\delta c(r)$:
\begin{align}
	R(r)=\frac{\sum_{|\mathbf{r}|=r} \sum_{\mathbf{r}^{\prime}} \delta c\left(\mathbf{r}^{\prime}\right) \delta c\left(\mathbf{r}+\mathbf{r}^{\prime}\right)}{\sum_{|\mathbf{r}|=r} \sum_{\mathbf{r}^{\prime}} \delta c\left(\mathbf{r}^{\prime}\right)^2},
\end{align}
where $\delta c(\mathbf{r}) = c(\mathbf{r}) - \bar{c}$ is the LCM fluctuation.
Similar to previous work, we determine the correlation length in the LCM as the distance at which the disorder-averaged autocorrelation function crosses zero \cite{Toma2020,Fuxiang2022}, $\bar{R}(\xi_r)=0$. The disorder-averaged autocorrelation function of the LCM in ground state, is shown in Fig.\ref{SM_Rspace_corfun}(b). 

Additionally, there is an alternative method to calculate correlation length. Another option is to use the nonlocal Chern marker, which is the off-diagonal elements of the Chern marker operator in Eq.\ref{LCMoperator}. This method has a direct relationship with the Fourier transform of the Berry curvature\cite{ChenWei2023sci}.

\section{Alternative characterization of the freeze-out time in momentum space}
\begin{figure}[!hbt]
	\centering
	\includegraphics[width=1\linewidth]{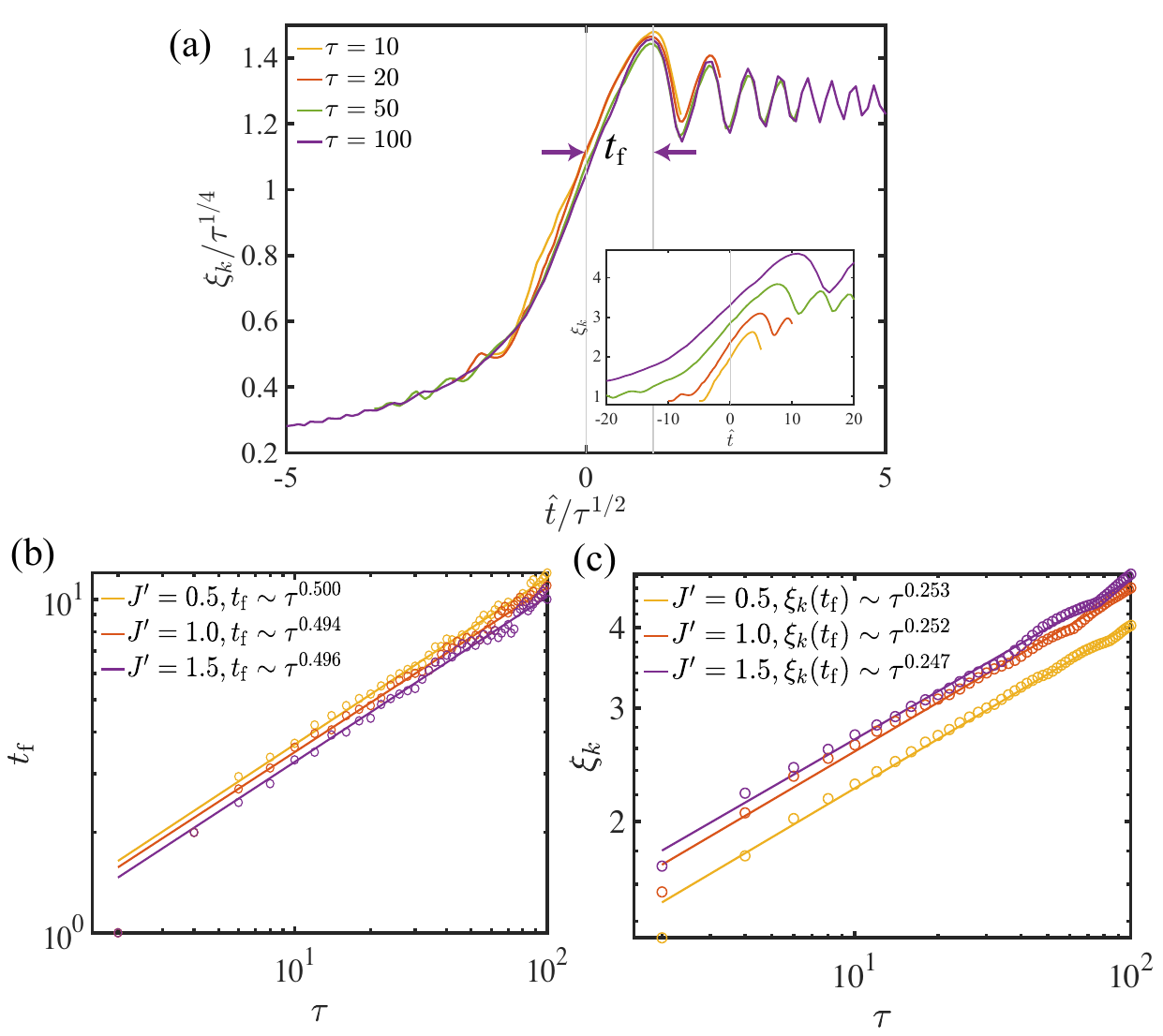}
	\caption{(a). The dynamics of correlation length in momentum space in a rescaled coordinate with $J^{\prime}=1$. Curves collapse near the transition point. The freeze-out time $t_\mathrm{f}$ can be determined by the first peak after the transition. The inset shows the same data in a non-rescaled coordinate. (b). Freeze-out time $t_\mathrm{f}$ has a power-law scaling with the quench time $t_\mathrm{f}\sim \tau^{1/2}$. (c). Correlation length at freeze-out time $t_\mathrm{f}$ shows a scaling behavior $\xi_k(t_\mathrm{f})\sim \tau^{1/4}$. }
	\label{SM_MaxCorlen_DelayTime}
\end{figure}
In the main text, we use the nonequilibrium Hall response to determine the freeze-out time during the quench in clean systems. The freeze-out time by definition characterizes the time when the system's relaxation rate matches the cooling rate. That is to say, it could be seen as a time delay when the system realizes the transition has occurred. So, the time delay of the trend of divergence of the correlation length can also be utilized to identify the freeze-out time $t_\mathrm{f}$. This method can be regarded as a supplement to the Hall response criterion and can be generalized to real space for disordered systems directly.

In Fig.\ref{SM_MaxCorlen_DelayTime}, we extract the freeze-out time $t_\mathrm{f}$ from the first peak after the transition. It shows the scaling of the freeze-out time $t_\mathrm{f}$ to the quench time $\tau$. And the unfrozen correlation length $\xi_k(t_\mathrm{f})$ also follows KZ's prediction.

\section{Alternative characterization of the freeze-out time in real space}
\begin{figure}[!hbt]
	\centering
	\includegraphics[width=1\linewidth]{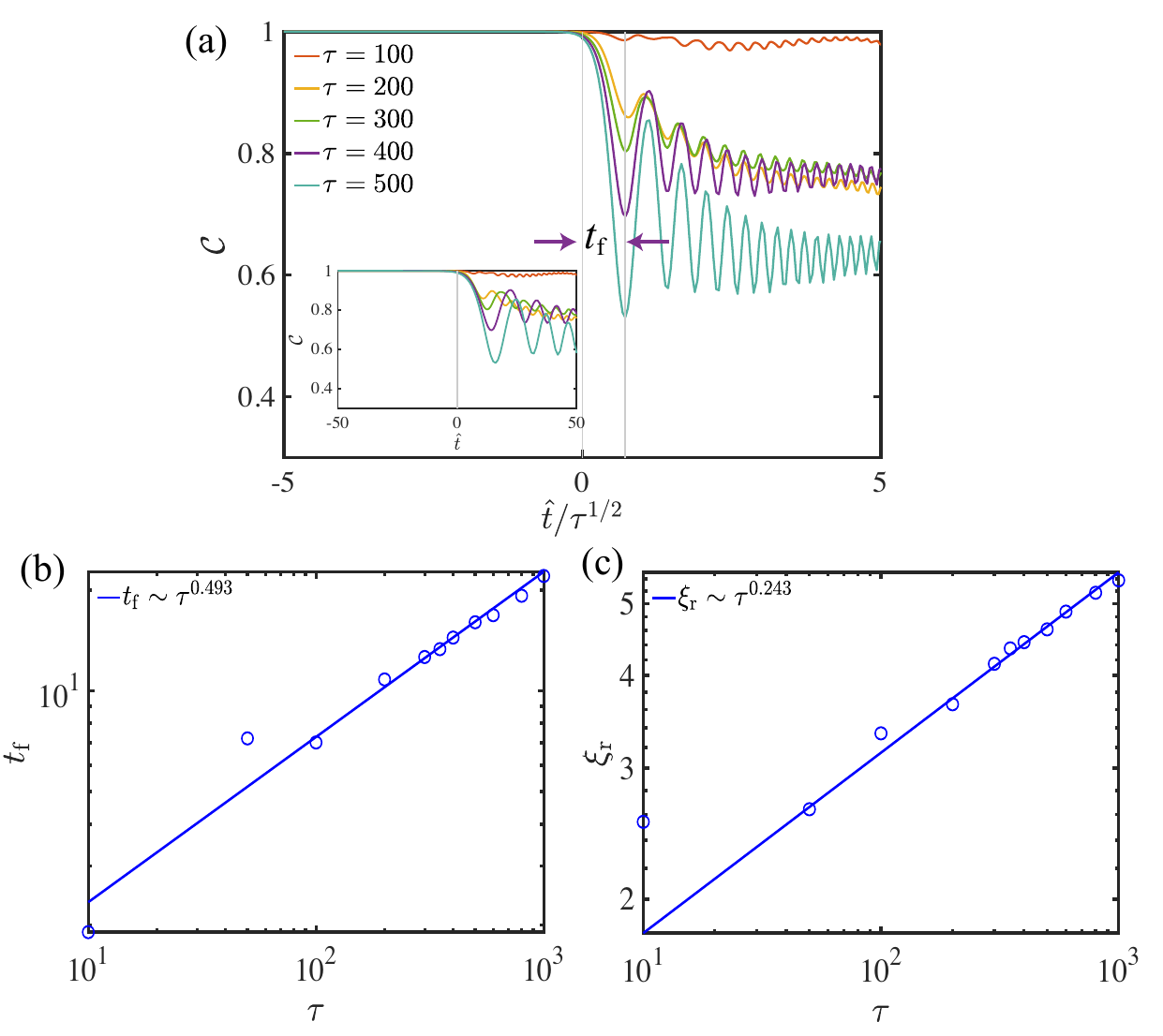}
	\caption{(a). The dynamics of the spatially averaged Chern marker $\mathcal{C}$ in a rescaled coordinate with size $N=50$. Curves show oscillation after the transition. The freeze-out time $t_\mathrm{f}$ can be determined by the first dip after the transition. The inset shows the same data in a non-rescaled coordinate. (b). Freeze-out time $t_\mathrm{f}$ has a power-law scaling with the quench time $t_\mathrm{f}\sim \tau^{1/2}$. (c). Correlation length at freeze-out time $t_\mathrm{f}$ shows a scaling behavior $\xi_r(t_\mathrm{f})\sim \tau^{1/4}$. }
	\label{SM_LCM_Rspace_DelayTime}
\end{figure}
In the main text, we determine the freeze-out time in disordered systems during the quench by the time delay at which the real-space correlation length reaches its maximum after the transition. For disordered systems, we find that the freeze-out time during the quench can also be determined by the time of the first dip in the spatially averaged Chern marker $\mathcal{C}$ after the transition.  

Spatially averaged Chern marker $\mathcal{C}$ is determined by the site-averaged and disorder-averaged LCM. It is quantized and equal to clean-band Chern number of the ground state in thermodynamic limit. But during the quench dynamics, $\mathcal{C}$ shows continuous oscillation after the transition point, and finally relaxes to the ground-state value, just similar as the Hall response (not shown within the time window of Fig.\ref{SM_LCM_Rspace_DelayTime}, because the relaxation is quite slow and may take time even longer than $\tau$). So, we can choose the first dip of the oscillation as a signature of the freeze-out time $t_\mathrm{f}$. Fig.\ref{SM_LCM_Rspace_DelayTime}
shows the extracted freeze-out time and corresponding unfrozen correlation lengths. It shows the scaling of the freeze-out time $t_\mathrm{f}$ to the quench time $\tau$. And the unfrozen correlation length $\xi_k(t_\mathrm{f})$ also follows KZ's prediction. 
\end{appendices}
%

\end{document}